\documentclass[twocolumn,twocolappendix]{aastex631}

\usepackage{amsmath,amssymb,bm}
\usepackage{graphicx}
\usepackage{booktabs}
\usepackage{xcolor}
\usepackage{mathrsfs}

\newcommand{\mdotbh}{\dot{M}_{\rm BH}}
\newcommand{\Msun}{M_\odot}
\newcommand{\mbh}{M_{\rm BH}}
\newcommand{\dMd}{\dot{M}_{\rm d}}
\newcommand{\tacc}{t_{\rm acc}}
\newcommand{\rg}{r_g}
\newcommand{\Ye}{Y_{\rm e}}
\newcommand{\Risco}{R_{\rm ISCO}}
\newcommand{\phibh}{\phi_{\rm BH}}

\newcommand{\Runb}{R_{\rm unb}}
\newcommand{\Rout}{R_{\rm out}}

\newcommand{\erg}{\mathrm{erg}}
\newcommand{\s}{\mathrm{s}}

\shorttitle{Collapsars Are Not Heavy $r$-Process Factories}
\shortauthors{Gottlieb et al.}

\begin{document}

\title{Collapsars Are Not Heavy $r$-Process Factories}

\author[0000-0003-3115-2456]{Ore Gottlieb}
\affil{Department of Physics and Kavli Institute for Astrophysics and Space Research, Massachusetts Institute of Technology, Cambridge, MA 02139, USA}
\email{ogot@mit.edu}

\author[0009-0005-2478-7631]{Danat Issa}
\affil{Department of Physics and Kavli Institute for Astrophysics and Space Research, Massachusetts Institute of Technology, Cambridge, MA 02139, USA}

\author[0000-0003-2982-0005]{Jonatan Jacquemin-Ide}
\affiliation{JILA, University of Colorado and National Institute of Standards and Technology, 440 UCB, Boulder, CO 80309-0440, USA}

\begin{abstract}

Collapsar disks are proposed sources of $r$-process elements, but neutron-rich disk ejecta require gas to be both neutronized in the disk and ejected before accretion. The magnetic field that enables ejection also thickens the disk and accelerates the inflow, lowering both the density and the time available for weak processing. To compute the resulting neutronization, we evolve the electron fraction of the inflowing gas, adopting at each radius the temperature that maximizes the net neutronization rate for the local density and instantaneous composition. This yields a lower limit on $Y_{\rm e}$ that is independent of the disk's thermal history. We adopt an optimistic unbinding radius of $R_{\rm unb}=10\,r_g$ for all magnetic fluxes capable of launching outflows. Combining this ejection condition with favorable assumptions of low black hole mass, ejection of midplane gas, and the upper end of the $\gamma$-ray burst (GRB) engine-power scale, which limits the accretion rate to $\dot M_{\rm BH}\lesssim0.4\,M_\odot\,{\rm s}^{-1}$, gives a fiducial lower bound $Y_{\rm e}\gtrsim0.33$, above the characteristic $Y_{\rm e}\approx 0.25$ threshold for heavy $r$-process production. Stronger magnetic flux, toward which jet-launching engines evolve, raises the lower bound to $Y_{\rm e}\gtrsim0.49$. We conclude that collapsar disk outflows cannot supply the heavy $r$-process yields invoked for a dominant enrichment channel. Only engines with unusually high accretion rates, potentially including exceptionally luminous GRBs, such as GRB~221009A, and direct-horizon collapsars, may produce heavy $r$-process ejecta.

\end{abstract}


\section{Introduction}
\label{sec:intro}
The rapid neutron-capture process ($r$-process) is responsible for roughly half of the abundances of nuclei heavier than iron. Producing the heavy $r$-process requires ejecta with a large neutron-to-seed ratio, set primarily by the electron fraction, entropy, expansion timescale, and subsequent neutrino irradiation \citep[e.g.,][]{Hoffman1997,Lippuner2015}; see \citet{Cowan1991,Cowan2021,Arnould2007,Qian2007,Thielemann2011,Thielemann2017,Kajino2019,Thielemann2026} for reviews. Neutron star (NS) mergers have long been predicted to synthesize $r$-process material \citep{Lattimer1974}, and the kilonova \citep{Li1998,Metzger2010,Metzger2020} associated with GW170817 \citep{Abbott2017} demonstrated that binary NS mergers are an important $r$-process source \citep[e.g.,][]{Cowperthwaite2017,Kasen2017,Pian2017,Smartt2017,Tanvir2017,Watson2019}.

Nevertheless, it remains unclear whether mergers alone account for the full diversity of $r$-process enrichment. The discovery of $r$-process-enhanced extremely metal-poor stars indicates that some enrichment occurred early, in low-metallicity environments \citep[e.g.,][]{Qian2000,Frebel2015,Ji2016,Ji2019,Roederer2016,Frebel2023,Simon2023}, motivating rare channels associated with massive-star deaths, whose short evolutionary timescales naturally couple them to early star formation \citep{Wehmeyer2015,Cote2019,Haynes2019,Kobayashi2020,Kobayashi2023,vandeVoort2020,Saleem2025}.

Collapsars -- rapidly rotating massive-star collapses that form proto-neutron stars followed by collapse into spinning black holes (BHs) surrounded by accretion disks \citep{Woosley1993,MacFadyen1999,MacFadyen2001,Woosley2006} -- have long been among the most promising massive-star alternatives \citep[e.g.,][]{Pruet2004,Surman2004,Surman2006,McLaughlin2005,Fujimoto2008,Janiuk2014}. If the disk becomes sufficiently dense and neutrino-cooled, electron degeneracy favors electron captures on protons, which can drive the gas neutron-rich \citep{Popham1999,Beloborodov2003,Pruet2003,Kohri2005,Chen2007}. The neutron-rich gas must then be ejected from the disk before it is accreted onto the BH.\footnote{An alternative neutronization mechanism outside the disk was proposed by \citet{Mumpower2025}, in which photohadronic reactions at the GRB jet head supply neutrons to the surrounding cocoon, potentially enabling $r$-process nucleosynthesis. However, their calculations rely on parameterized conditions, and it remains unclear whether this mechanism produces substantial heavy-element yields in a self-consistently evolving collapsar.}

\citet{Siegel2019} performed general-relativistic magnetohydrodynamic (GRMHD) simulations to follow the interplay between outflow dynamics and nucleosynthesis, finding substantial $r$-process production in disk outflows. Similarly, accreting proto-magnetars in the pre-BH phase can also eject neutron-rich disk material through interaction with the stellar magnetosphere \citep{Prasanna2026}. However, those calculations began from an idealized, pre-assembled torus and did not include the ram pressure of the stellar envelope, self-consistent disk formation, or the continued interaction between fallback and the disk. The robustness of $r$-process production in such idealized tori was questioned by simulations with detailed neutrino transport \citep{Miller2020}.

The core difficulty became clearer in recent global hydrodynamic collapsar simulations \citep{Just2022,Fujibayashi2023,Dean2024a,Dean2024b}. These studies found that, although neutron-rich gas can form in the inner disk during the neutrino-dominated accretion flow (NDAF) phase, it is accreted before being ejected. These works suggested that magnetic fields may be essential, as sufficiently strong magnetic stresses could more efficiently launch neutron-rich material before it is swallowed by the BH. Indeed, \citet{Metzger2008} showed that a strong poloidal field allows NDAF outflows to retain their neutron excess, provided that the gas is neutron-rich where it is launched.

The first 3D global collapsar simulations with both magnetic fields and neutrino transport found that this tension is fundamental \citep{Issa2025}, and axisymmetric simulations similarly find that $r$-process elements are only weakly synthesized once a jet is launched \citep{Shibata2025}. The two requirements for robust $r$-process outflows are not independent. First, the gas in the disk must cool and neutronize before being advected onto the BH, which requires neutrino cooling and weak-interaction timescales shorter than the inflow time. Second, the resulting neutron-rich gas must be unbound by magnetic stresses before it is swallowed. Magnetic fields help satisfy the second requirement, but they also enhance angular-momentum transport, shorten the inflow time, provide magnetic and thermal support, and thicken the disk. These effects suppress the neutrino-cooled, mildly degenerate state required for low electron fraction $\Ye$ material, thereby making neutronization more difficult. Thus, the same magnetic stresses needed to eject neutron-rich gas can also erase the conditions required to produce it.

In the simulation suite of \citet{Issa2025}, neutron-rich outflows appeared only in the most extreme high-accretion-rate progenitors, where the disk remained neutrino cooled despite strong magnetic stresses. Therefore, these models do not demonstrate that collapsars associated with the observed long $\gamma$-ray burst (GRB) population are $r$-process factories. Such engines produce outflow power well above that inferred for GRBs or similarly energetic engine-driven supernova-like transients. In this paper, we address whether collapsar disk outflows can satisfy the neutronization and ejection requirements while remaining consistent with the observed GRB engine power scale.

Two loopholes remain in that engine-power scale. The first is a quasi-steady compromise between the weakly magnetized, efficiently cooled standard-and-normal-evolution disk \citep[SANE;][]{Narayan2012} limit and the strongly magnetized magnetically arrested disk \citep[MAD;][]{Narayan2003,Tchekhovskoy2011} limit. In this picture, an intermediate state resembling a rocking accretion disk \citep[RAD;][]{Lalakos2024} could retain enough coherent magnetic flux to eject material while still allowing the disk to cool and neutronize. The second is temporal separation: the disk first neutronizes while weakly magnetized, and only later, as magnetic flux accumulates near the BH and generates a coherent, large-scale field via the dynamo process \citep[e.g.,][]{Jacquemin-Ide2024b,Chan2026}, does it eject the gas that was neutronized beforehand. These two scenarios define the remaining ways in which collapsars associated with observed GRBs could still act as heavy $r$-process factories.

\begin{figure*}
    \centering
    \includegraphics[width=1\linewidth]{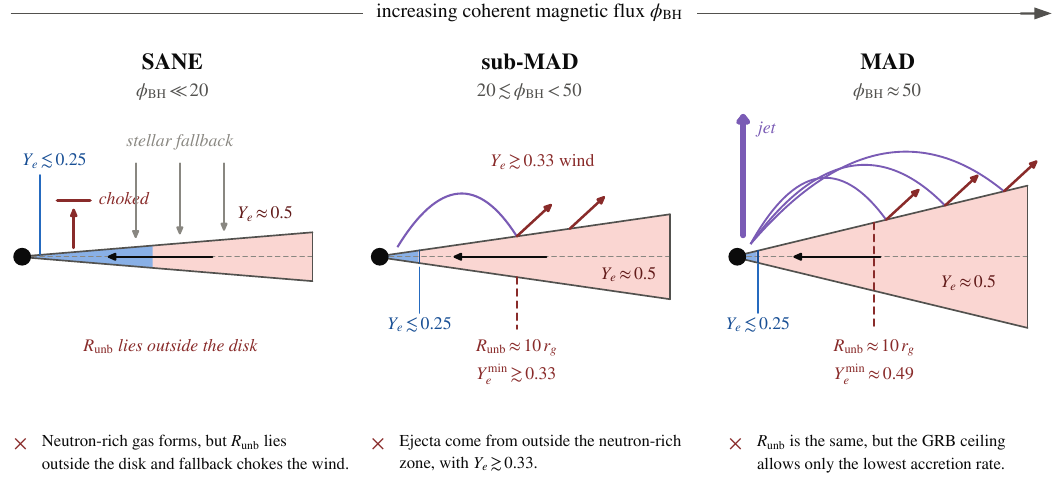}
    \caption{Disk neutronization and ejection regimes of dimensionless magnetic flux, $\phibh$. Colors show the electron fraction (blue for $\Ye\lesssim0.25$, and red for $\Ye\approx 0.5$). \emph{SANE}: efficient neutrino cooling makes a dense, neutron-rich midplane, but winds cannot overcome fallback ram pressure and $\Runb$ lies outside the disk. \emph{sub-MAD}: finite coherent flux threading the disk (purple bundles) launches winds, but the same magnetic support shortens the inflow and lowers the density, such that even the optimized bound at $\phibh=20$ stays above $\Ye\approx0.33$ down to $10\,r_g$. \emph{MAD}: fields readily drive jets and disk winds, but the stronger stresses thicken the disk and erase the degenerate, neutrino-cooled state. The accompanying reduction in the allowed accretion rate further suppresses the integrated neutronization. On the GRB ceiling, $\Ye^{\min}(\Runb)$ rises from 0.33 at $\phibh=20$ to 0.49 at $\phibh=50$. Therefore, none of the states ejects gas with $\Ye\lesssim0.25$.}
    \label{fig:sketch}
\end{figure*}

We test these possibilities by computing the lowest electron fraction that inflowing gas can reach before ejection. Magnetic ejection becomes dynamically important only outside an effective unbinding radius, $\Runb$. Thus, disk-wind $r$-process production requires neutron-rich gas to be present at $\Runb$. Fig.~\ref{fig:sketch} illustrates how collapsars associated with observed GRBs fail the above condition. In weakly magnetized disks, gas can cool and neutronize but remains bound and is accreted. In more strongly magnetized disks, magnetic stresses can unbind gas from $\sim10\,r_g$, but the same stresses thicken the disk and shorten the inflow time, such that the gas reaches that radius only weakly neutronized.

The structure of this paper is as follows. In \S\ref{sec:grb_norm}, we use the observed long GRB luminosity scale to estimate the accretion rates and magnetic powers relevant for collapsars. In \S\ref{sec:flux_sequence}, we test whether gas in such disks can neutronize before reaching the adopted launching radius, considering the SANE--MAD magnetization sequence and the possibility of gradual magnetization. In \S\ref{sec:others}, we discuss the implications for luminous GRBs and rare high-accretion-rate collapsars. We summarize our conclusions in \S\ref{sec:conclusions}.

\section{GRB Observational constraints}
\label{sec:grb_norm}

The power of magnetically driven Blandford-Znajek \citep{Blandford&Znajek1977} outflows can be written as \citep{Gottlieb2023e}
\begin{equation}\label{eq:BZ}
    P_j = \eta_a\eta_\phi\mdotbh c^2,
\end{equation}
where $a$ is the BH Kerr parameter, $\mdotbh$ is the mass accretion rate onto the BH, $\eta_a$ parameterizes the efficiency with which the BH rotational energy is extracted \citep{Lowell2024},
\begin{equation}\label{eq:eta_a}
\eta_a \approx \left(1.06a^4+0.395a^2\right),
\end{equation}
and $\eta_\phi$ parameterizes the fraction of the MAD-level dimensionless magnetic flux, $\phibh^{\rm MAD}\approx 50$, threading the BH \citep{Tchekhovskoy2011},
\begin{equation}
    \eta_\phi \approx \left(\frac{\phibh}{\phibh^{\rm MAD}}\right)^2,
\end{equation}
where
\begin{equation}\label{eq:phibh}
    \phibh=\frac{\Phi}{r_g\sqrt{\mdotbh c}},
\end{equation}
with $\Phi$ being the poloidal magnetic flux threading the BH, and $r_g=G\mbh/c^2$ is the BH gravitational radius.

The mass accretion rate, $\mdotbh$, in Eq.~\eqref{eq:BZ} is constrained by the observed luminosities of long GRB jets. For jets with $\theta_j\approx0.1$, the beaming-corrected prompt $\gamma$-ray \emph{peak} luminosities, $L_j=L_{\rm iso}\theta_j^2/2$, are typically
$10^{49}\,\erg\,\s^{-1}\lesssim L_j\lesssim 10^{51}\,\erg\,\s^{-1}$ \citep{Wanderman2010}. Adopting a fiducial prompt radiative efficiency $\eta_\gamma\simeq 0.2$ \citep{Zhang2007,Beniamini2016,Aksulu2022}, the total jet power is $P_j\sim L_j/\eta_\gamma\lesssim5\times10^{51}\,\erg\,\s^{-1}$. Eq.~\eqref{eq:BZ} then gives a maximum mass accretion rate
\begin{equation}
\mdotbh\lesssim 10^{-2.5}\,\frac{\Msun}{\s}
 \left(\frac{P_j}{5\times10^{51}\,\erg\,\s^{-1}}\right)
 \left(\eta_a\eta_\phi\right)^{-1},
 \label{eq:macc_general}
\end{equation}
over the observed GRB duration. Thus, high accretion rates are possible only if the launching efficiency $\eta_a\eta_\phi$ is well below unity.

The flux factor, however, cannot be arbitrarily small as jets emerge from the ergosphere only once $\phibh\sim20$ \citep{Issa2025,Lalakos2025}, corresponding to $\eta_\phi\gtrsim 0.16$. The remaining suppression must therefore come largely from the spin factor. Collapsar BHs are moderately spinning with $a\approx 0.3$ \citep{Gottlieb2023,Jacquemin-Ide2024}, for which Eq.~\eqref{eq:eta_a} gives $\eta_a\sim0.05$. Substantially larger spins would produce jet luminosities well above the observed GRB range, as found in simulations initialized with rapidly spinning BHs \citep{Gottlieb2022c,Shibata2024}. Lower spins are not directly excluded by the GRB luminosity distribution, because their lower BZ efficiency can be compensated by a larger accretion rate to produce the same jet power. However, jets powered by slower-spinning BHs would be less powerful than the confining accretion flow \citep{McKinney2012,Narayan2012,Tchekhovskoy2012,Bopp2025}, making them more susceptible to choking before breakout. While such low-spin, high $\dot M_{\rm BH}$ engines could then remain hidden while hosting neutron-rich disk outflows, reconciling such a population with GRB observations would require a systematic anti-correlation between BH spin and accretion rate. Otherwise, the same high $\dot M_{\rm BH}$ population should also include some BHs with higher spins whose jets would again conflict with the observed GRB luminosity distribution. Such a tuned natal anti-correlation is not a generic prediction of current progenitor models \citep{Renzo2026}.

The value of $\phibh$, and hence $\eta_\phi$, is expected to grow during the collapse, for example if the magnetic flux threading the BH is generated or amplified by dynamo processes \citep{Shibata2025,Chan2026}. Once a jet is launched, mass accretion suppression and ongoing disk dynamo would further increase $\phibh$ until the flow approaches the MAD state, where $\eta_\phi\sim 1$. Thus, although the MAD accretion rate during most of the observed GRB emission is limited to $\mdotbh\lesssim 0.1\,\Msun\,\s^{-1}$, the transition toward MAD can temporarily allow higher accretion rates. Using Eq.~\eqref{eq:macc_general}, and plugging in $\phibh\approx 20$ as the minimum magnetic flux for sustained jet launching, the maximum mass accretion rate is
\begin{equation}\small
\dot{M}_{\rm BH, max}\approx 0.4\,\frac{\Msun}{\s}\left(\frac{a}{0.3}\frac{\phibh}{20}\right)^{-2}\left(\frac{P_j}{5\times10^{51}\,\erg\,\s^{-1}}\right),
 \label{eq:Mdotbh}
\end{equation}
where the normalization uses Eq.~\eqref{eq:eta_a} at $a=0.3$, while the spin dependency follows the low-spin scaling, $\eta_a\sim a^2$. We adopt this value as an optimistic upper limit for collapsars. In \S\ref{sec:others}, we return to the possibility of atypical, much higher-$\mdotbh$ progenitors.

\section{Failure across the magnetic-flux sequence}
\label{sec:flux_sequence}

In this section, we examine the collapsar problem across the magnetic-flux sequence. We first consider the weak-flux limit, where magnetic ejection is ineffective and neutron-rich gas is predominantly accreted. We then consider sub-MAD/MAD disks capable of launching winds, assigning all of them the same favorable unbinding radius, $R_{\rm unb}=10\,r_g$, and computing the minimum electron fraction there. Finally, we consider whether temporal separation between neutronization and magnetic-flux accumulation can evade this constraint.

\subsection{Weak-flux/SANE limit}
\label{sec:sane}
The weak-flux/SANE limit corresponds to $\phibh < 20$, below which no jet emerges. Without a jet, the disk launches magnetically driven winds only once it becomes MAD \citep{Issa2025}, so any outflow in this regime must be driven hydrodynamically. This regime is also the most favorable for neutronization since the residence time is long and the density is high. Global hydrodynamic calculations with prescribed viscous angular momentum transport provide a useful proxy for the weak coherent flux limit. These simulations find that the low $\Ye$ material formed in the inner neutrino-cooled disk is predominantly accreted rather than ejected \citep{Just2022,Fujibayashi2023,Dean2024a,Dean2024b}. Outflows emerge only after the disk has become less efficiently cooled and correspondingly less neutron rich. As a result, the ejecta contain negligible abundances of $r$-process.

The global calculations suggest that hydrodynamic disks inhibit neutron-rich ejecta through the combination of (i) limited available thermal energy in the efficiently cooled inner disk to drive outflows, and (ii) ram pressure of the fallback material:

(i) In the NDAF regime, the neutrino cooling time is much shorter than the accretion time,
\begin{equation}    \label{eq:weak_acc}
    \frac{t_\nu}{\tacc}\ll1.
\end{equation}
Therefore, most of the thermal energy generated by dissipation is radiated before it can be advected, leaving only a small residual thermal energy reservoir to launch a hydrodynamic wind from the NDAF region\footnote{Nonlocal viscous energy transport and nuclear recombination may provide additional power. Neutrinos emitted by the dense disk can also heat its atmosphere and contribute to mass loss during the NDAF phase. However, global collapsar calculations that include neutrino transport find that these outflows are insufficiently neutron rich \citep{Just2022}.}. Thus, the mechanical power of such a wind satisfies
\begin{equation}\label{eq:Eavail}
    P_{\rm hyd} \lesssim \frac{t_\nu}{\tacc}\, \dMd e_{\rm bind},  
\end{equation}
where $\dMd$ is the local inward accretion rate through the disk, and the characteristic binding energy per unit mass of Keplerian disk gas is 
\begin{equation}\label{eq:ebind}
    e_{\rm bind}(r)\sim \frac{G\mbh}{2r}. 
\end{equation}
To unbind a hydrodynamic wind with mass flux $\dot{M}_w$, one must supply at least the local binding energy,
\begin{equation}\label{eq:Ereq}
    P_{\rm hyd}\gtrsim\dot{M}_we_{\rm bind}.
\end{equation} Combining Eqs.~\eqref{eq:weak_acc}, \eqref{eq:Eavail} and~\eqref{eq:Ereq} gives
\begin{equation}\label{eq:Mwind_hydro_bound} 
    \frac{\dot{M}_w}{\dMd}\lesssim\frac{t_\nu}{\tacc}\ll1.
\end{equation}

(ii) Unlike a post-merger disk, a collapsar disk remains embedded in and continuously replenished by the infalling star. The disk wind must push against the ram pressure of the freefalling gas with velocity $v_{\rm ff}$. For a broad hydrodynamic wind with mass flux $\dot M_w$ and velocity $v_w$, we compare the radial momentum fluxes per solid angle to find
\begin{equation}
\dot{ M}_wv_w\gtrsim\dot{M_{\rm ff}}v_{\rm ff}.
\label{eq:ram_pressure_condition}
\end{equation}
Approximating $\dot{M}_{\rm ff}\sim\dMd$, this gives
\begin{equation}\label{eq:Mdot_ratio}
\frac{\dot M_w}{\dMd}\gtrsim\frac{v_{\rm ff}}{v_w},
\end{equation}
but this condition cannot be satisfied since Eq.~\eqref{eq:Mwind_hydro_bound} implies that the L.H.S. is very small whereas $v_{\rm w}$ must be comparable to $v_{\rm ff}$ in order for disk outflows to emerge.

These arguments do not exclude hydrodynamic outflows, but disfavor the massive, sufficiently low-$\Ye$ wind required for collapsars to dominate $r$-process enrichment. Coherent magnetic stresses provide a different route to angular-momentum removal and ejection.

\subsection{Sub-MAD/MAD regime}\label{sec:mad}
As sub-MAD/MAD disks can launch outflows, the question is no longer whether ejection occurs but whether it takes place in neutron-rich regions. We adopt an optimistic inner launching radius and compute the lowest electron fraction attainable before the gas reaches it.

We define $\Runb$ as an effective magnetic unbinding radius: the radius at which the available magnetic work per loaded mass becomes comparable to the local binding energy of the gas. We adopt a common unbinding radius,
\begin{equation}
R_{\rm unb}=10\,r_g,\qquad20\lesssim\phi_{\rm BH}\lesssim50.
\end{equation}
This choice is motivated by the global MAD simulations of \citet{Lalakos2025}, whose mass-loss profiles show dynamically important outflows beginning at $\approx10\,r_g$. These simulations are non-radiative, allowing the gas to retain dissipated heat that can assist ejection. Neutrino cooling removes part of this thermal support, while stellar fallback further resists disk winds. We therefore adopt $10\,r_g$ as an optimistic launching radius for collapsar outflows.
Assigning the same radius to sub-MAD disks grants them the favorable ejection condition of the MAD state. This allows the gas to accumulate weak processing down to $10\,r_g$ at every magnetic flux capable of launching a wind. If appreciable mass loss instead begins at larger radii, the gas has less time to neutronize along the same inflow trajectory, and the resulting composition is less neutron rich.

The conventional NDAF ignition condition, obtained by comparing viscous heating with optically thin neutrino cooling by electron and positron captures on free nucleons, has commonly been used to estimate the neutron richness in disks \citep{Beloborodov2003,Chen2007,Metzger2008}. However, efficient cooling alone is not sufficient for self-neutronization as the gas must also reach a sufficiently neutron-rich thermodynamic state, and weak interactions must act faster than the inflow time. Indeed, \citet{Agarwal2026} found that for some massive BHs, the ignition condition can be satisfied while self-neutronization fails. Moreover, the agreement between the conventional ignition condition and the composition found in previous simulations pertains to an MRI-turbulent, effectively viscous accretion regime, rather than to the coherent high-flux sub-MAD/MAD flows required for mass ejection, in which the disk dynamics is modified by large-scale magnetic torques \citep{Blandford&Payne1982}. These torques do not obey the local viscous-heating closure entering the standard ignition estimate. In the discussion below, we therefore estimate the composition directly.

\begin{figure*}
    \centering
    \includegraphics[width=1\linewidth]{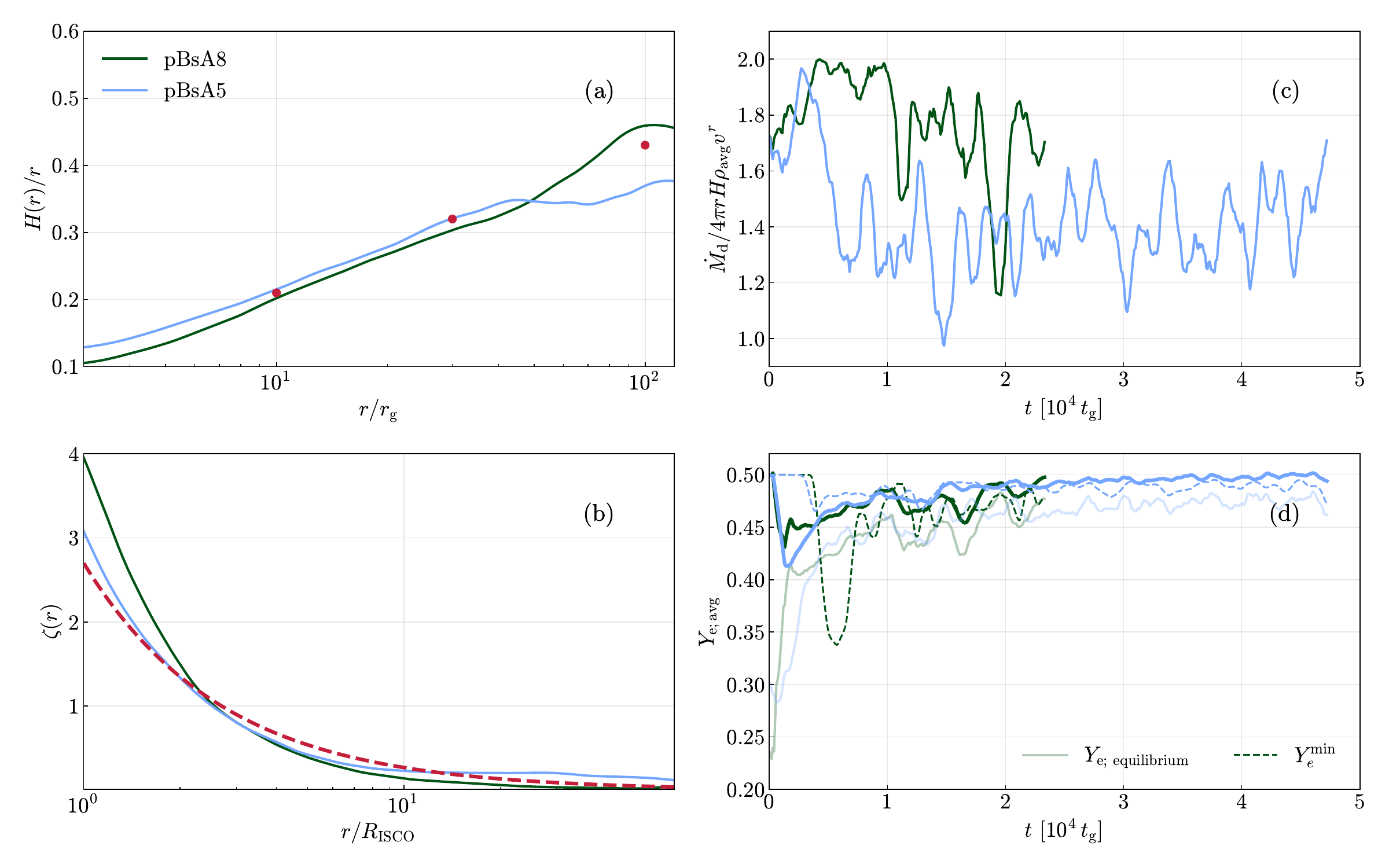}
    \caption{Comparison of analytic fits with two GRMHD simulations, $\rm pBsA8$ ($a=0.8$) and $\rm pBsA5$ ($a=0.5$). The models are based on model pBs from \citet{Issa2025}, with the indicated BH spins and a corrected neutrino absorption scheme.
    {\bf (a)} Time-averaged $H/r$ profiles, and representative values of our fit: $H_0(10\,r_g)=0.21\,r$, $H_0(30\,r_g)=0.32\,r$, and $H_0(100\,r_g)=0.43\,r$ (red markers).
    {\bf (b)} Time-averaged $\zeta$ profiles, and our fit to the numerical profile at $r>10\,r_g$ in Eq.~\eqref{eq:zeta_profile} (red dashed line).
    {\bf (c)} Time evolution in $t_g\equiv \rg/c$ of Eq.~\eqref{eq:mdot_disk} at $r=10\,\rg$.
    {\bf (d)} Time evolution of different characteristic $\Ye$ at $r=10\,\rg$: instantaneous density-weighted average (dark solid lines), equilibrium value (faint solid lines), and the analytic estimate for the minimum value from Eq.~\eqref{eq:Ye_analytic} (dashed lines). The $\Ye$ comparison demonstrates that $\Ye \gtrsim Y_{\rm e;\,equilibrium}$ at all times, and that the analytic estimate provides a reasonable lower bound.}
    \label{fig:numerical}
\end{figure*}

Local turbulent transport has inward velocity $v_{\rm in}\sim\alpha_{\rm visc}(H/r)^2v_K$, where $v_K$ is the Keplerian velocity and $H$ is the disk scale height. In contrast, coherent magnetic torques scale linearly with $H/r$ and can drive substantially faster accretion in thinner disks \citep{ferreira_magnetized_1995,casse_magnetized_2000,Petrucci2010,jacquemin-ide_magnetically-driven_2019}. Numerical simulations of MADs have found that angular-momentum transport is dominated by coherent large-scale torques for $H/r\lesssim0.2$ \citep{scepi_2024_mads}, while even in geometrically thick adiabatic MADs, these torques can contribute comparably to turbulent transport \citep{manikantan_2024_mads}. In collapsars, $H/r$ spans a broad range, from $\sim0.1$ near the innermost stable circular orbit, $R_{\rm ISCO}$, to $\sim0.5$ at $r\sim100\,r_g$ [see e.g., Fig.~\ref{fig:numerical}(a)]. A fully general treatment would therefore need to account for both transport channels, with coherent torques dominating close to the BH and turbulent stresses becoming more important at larger radii. For simplicity, we retain only the $H/r$ dependence expected for coherent large-scale torques and introduce a radial function, $\zeta(r)$, to capture the behavior farther from the BH.
Thus, we write
\begin{equation}
\label{eq:magnetic_inflow}
 v_{\rm in}(r)=\zeta(r)\frac{H(r)}{r}v_K(r),
\end{equation}
where $\zeta$ is a kinematic parameter. We adopt a quasi-steady flow with ignition-friendly low BH mass of $\mbh=3\,\Msun$.

If coherent torques dominate the accretion velocity \citep{ferreira_magnetized_1995,casse_magnetized_2000,jacquemin-ide_magnetically-driven_2019}, then $\zeta$ is determined by 
\begin{equation}
    \frac{v_{\rm in}}{v_K} \sim \frac{H}{r}\frac{B_{\hat{\theta}}B_{\hat{\varphi}}}{P}\sim\zeta\frac{H}{r};\qquad\zeta\sim\frac{B_{\hat{\theta}}B_{\hat{\varphi}}}{P},
\end{equation}
where $P$ is the thermal pressure, and $B_{\hat{\theta}}$ and $B_{\hat{\varphi}}$ are the coherent components of latitudinal and azimuthal fields. In MAD, it is expected that $B_{\hat{\theta}}^2\lesssim P$ and $B_{\hat{\varphi}}^2\sim P $. Therefore, we expect $\zeta\sim1$ as long as the disk is magnetically dominated and coherent torques dominate accretion. Fig.~\ref{fig:numerical}(a),(b) depicts our fits to the numerical data of $H/r$ and $\zeta$ in neutrino-cooled disks in GRMHD simulations from \citet{Issa2025}. For the inflow normalization in Fig.~\ref{fig:numerical}(b), we fit\footnote{This radial dependency of $\zeta$ can be understood as a consequence of the accretion disk becoming thicker and therefore thermal pressure and turbulent transport becoming stronger than the magnetic fields farther out.}
\begin{equation}
\label{eq:zeta_profile}
 \zeta_0(r)=0.27\left(\frac{r}{10\,\Risco}\right)^{-1}.
\end{equation}
The corresponding accretion timescale is
\begin{equation}
\label{eq:tacc}
\begin{split}
 \tacc(r)&\simeq{}1.5\,{\rm ms}
 \left(\frac{\mbh}{3\,\Msun}\right)
 \left(\frac{H(r)/r}{0.23}\right)^{-1}\times\\
 &\left(\frac{\Risco(a)}{\Risco(a=0.3)}\right)^{-1}
 \left(\frac{\zeta}{\zeta_0}\right)^{-1}
 \left(\frac{r}{10\,r_g}\right)^{\frac{5}{2}}.
\end{split}
\end{equation}

To estimate the radial density profile, we write the vertically integrated disk mass conservation
\begin{equation}
\label{eq:mdot_disk}
 \dMd=2\pi r\Sigma v_{\rm in}
 \simeq4\pi rH\rho v_{\rm in},
\end{equation}
where $\Sigma\simeq2\rho H$. Then
\begin{equation}
\label{eq:rho}
\begin{split}
 \rho(r)&\simeq{}
 \frac{\dMd r^{-3/2}}
 {4\pi\zeta(r)[H(r)/r]^2\sqrt{G\mbh}}\\
 &\simeq{}4.8\times10^9\,{\rm g\,cm^{-3}}
 \left(\frac{\dMd}{0.4\,\Msun\,{\rm s}^{-1}}\right)
 \left(\frac{\mbh}{3\,\Msun}\right)^{-2}\times\\
 &\left(\frac{H(r)/r}{0.23}\right)^{-2}
 \left(\frac{\Risco(a)}{\Risco(a=0.3)}\right)^{-1}
 \left(\frac{\zeta}{\zeta_0}\right)^{-1}
 \left(\frac{r}{10\,r_g}\right)^{-\frac{1}{2}}.
\end{split}
\end{equation}
Fig.~\ref{fig:numerical}(c) demonstrates that the mass-flux estimate in Eq.~\eqref{eq:mdot_disk} agrees with the GRMHD collapsar simulations to within a factor of two.

We estimate the finite net neutronization rate using the electron and positron capture integrals of \citet[][Eqs.~6--7]{Beloborodov2003}. We assume free, nondegenerate nucleons and negligible neutrino absorption (see Appendix~\ref{app:neutrinos}, Fig.~\ref{fig:neutrinosphere}), while retaining arbitrary electron degeneracy through the chemical potential determined by charge neutrality. The net rate is
\begin{equation}
\label{eq:net_neutronization}
 \Gamma(\rho,T,\Ye)\equiv-\frac{d\Ye}{dt}
 =\Ye\lambda_{e^-p}-(1-\Ye)\lambda_{e^+n},
\end{equation}
where the capture coefficients $\lambda$ are rates per target nucleon. We neglect the formation of $\alpha$ particles when evaluating the weak-interaction rates, which makes our neutronization bound conservative because it overestimates the free nucleon fraction (see Appendix~\ref{app:alpha}). At each density and instantaneous composition, we find the temperature that maximizes the instantaneous net conversion rate (Appendix~\ref{app:tstar}),
\begin{equation}
\label{eq:neutronization_bound}
\frac{d\Ye}{dr}=\frac{\Gamma_\star[\rho(r),\Ye(r)]}{v_{\rm in}(r)};
\quad \Gamma_\star(\rho,\Ye)=\max_{T\geq0}\Gamma(\rho,T,\Ye).
\end{equation}
The numerical integration proceeds from $\Rout$ with $\Ye(\Rout)=0.5$ inward to $\Runb=10\,r_g$. Beyond $\Rout\approx100\,r_g$, the heat per nucleon of an advective disk, $\frac{3}{2}S\,(r_g/r)\,m_uc^2\approx7\,{\rm MeV}\,(100\,r_g/r)$, where $S\simeq1/2$ accounts for the zero-torque inner boundary \citep[][Eqs.~36--41]{Beloborodov2003}, falls below the $7.07\,$MeV binding energy per nucleon of $\alpha$ particles, which do not capture electrons. Starting instead at $3000\,r_g$ in nuclear statistical equilibrium lowers $\Ye^{\rm min}$ by at most $0.02$ along the ceiling (see Appendix~\ref{app:rout}).

\begin{figure}
    \centering
    \includegraphics[width=1\linewidth]{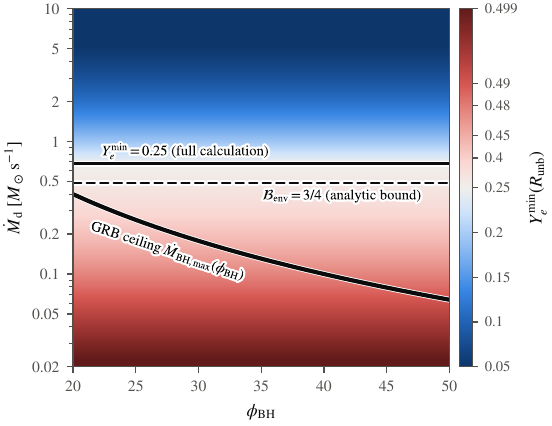}
    \caption{Minimum $\Ye(\Runb)$ in $(\dMd,\phibh)$. The thin solid (dashed) line shows the $\Ye=0.25$ contour using the numerical (analytic) estimate. The thick line delineates the ceiling of the GRB population: $\dot M_{\rm BH,max}=0.4\,\Msun\,{\rm s^{-1}}\,(20/\phibh)^2$, for $P_j=5\times10^{51}\,{\rm erg\,s^{-1}}$, $a=0.3$, and $\mbh=3\,\Msun$. Along the GRB ceiling, $\Ye^{\min}\gtrsim0.33$ for any value of $\phibh$.
    }
    \label{fig:dependencies}
\end{figure}

Taking $\mdotbh\simeq\dMd$ inside $r < \Rout$ \citep[see e.g.,][]{Lalakos2025}, Fig.~\ref{fig:dependencies} shows the minimum attainable $\Ye$ as a function of $\dMd$ and $\phibh$, while fixing the other parameters to their canonical values. The $\Ye^{\min}=0.25$ contour (thin solid line), adopted as the threshold for heavy $r$-process production \citep[e.g.,][]{Lippuner2015}, remains above the GRB ceiling (thick line). Along this ceiling, the minimum attainable electron fraction increases from $\Ye^{\min}(\phibh=20)\simeq0.33$ to $\Ye^{\min}(\phibh=50)\simeq0.49$. In Appendix~\ref{app:rout}, we show that allowing additional weak processing at larger $\Rout$ lowers the most favorable value to $\Ye^{\min}\sim0.31$--$0.33$. Furthermore, the most favorable state, $\phi_{\rm BH}\simeq20$, is likely to be transient. Continued accumulation of coherent magnetic flux and suppression of accretion tend to drive the flow toward MAD, which lowers the permitted accretion rate and suppresses neutronization. The most favorable conditions therefore likely operate only during a limited sub-MAD phase.

To understand the dependencies of $\Ye$ on the collapsar disk properties, we additionally provide an analytic bound. Using the mildly degenerate expansion and charge-neutrality relation of \citet[][Eqs.~9--10 and 13]{Beloborodov2003}, the net rate is
\begin{equation}
\label{eq:rate_expansion}
 \Gamma(\rho,T,0.5)\simeq
 \frac{7\pi^4K}{60}\left(D\theta^2-Q\theta^4\right),
\end{equation}
where $\theta=kT/(m_ec^2)$, $D=3\lambdabar_e^3\rho/m_p$, $\lambdabar_e=\hbar/(m_ec)$, $Q=(m_n-m_p)/m_e=2.531$, and $K=6.5\times10^{-4}\,{\rm s}^{-1}$. Maximizing over temperature gives $\theta^2=D/(2Q)$ and $\Gamma_\star(\rho,0.5)\simeq7.8\,(\rho/10^9\,{\rm g\,cm^{-3}})^2\,{\rm s}^{-1}$. Comparison with the full capture integrals gives a conservative rate upper limit\footnote{The expansion in Eq.~\eqref{eq:rate_expansion} assumes $\mu_e < \theta$, where $\mu_e$ is the electron chemical potential, and $\theta > Q+1$. Therefore $\Gamma_{\rm env}$ overestimates the true rate at low densities, which justifies it as an upper limit estimate.},
\begin{equation}
\label{eq:rate_envelope}
 \Gamma_\star(\rho,0.5)\lesssim\Gamma_{\rm env}(\rho)
 \equiv8\,\left(\frac{\rho}{10^9\,{\rm g\,cm^{-3}}}\right)^2\,{\rm s}^{-1}.
\end{equation}

At fixed temperature, the states $(\rho,\Ye)$ and $(2\Ye\rho,0.5)$ have identical per-target capture rates since they scale with $\rho\Ye$. Hence,
\begin{equation}
 \Gamma(2\Ye\rho,T,0.5)
 =\frac{1}{2}
 \left(\lambda_{e^-p}-\lambda_{e^+n}\right),
\end{equation}
where the rates on the R.H.S. are those evaluated at $(\rho,T,\Ye)$. Multiplying by $2\Ye$ and comparing with Eq.~\eqref{eq:net_neutronization} gives
\begin{equation}
 \Gamma(\rho,T,\Ye)
 =2\Ye\Gamma(2\Ye\rho,T,0.5)
 -(1-2\Ye)\lambda_{e^+n}.
\end{equation}
Maximizing over temperature therefore gives an upper bound on the net neutronization rate,
\begin{equation}\label{eq:composition_rate_bound}
\Gamma_\star(\rho,\Ye)\leq2\Ye\,\Gamma_\star(2\Ye\rho,0.5)\lesssim(2\Ye)^3\Gamma_{\rm env}(\rho),
\end{equation}
where in the R.H.S. we applied Eq.~\eqref{eq:rate_envelope} at the rescaled density, and assumed $\Ye<0.5$.

We define the integrated rate bound
\begin{equation}
\label{eq:Benv}
 \mathcal B_{\rm env}\equiv
 \int_{\Runb}^{\Rout}
 \frac{\Gamma_{\rm env}[\rho(r)]}{v_{\rm in}(r)}\,dr.
\end{equation}
Using Eq.~\eqref{eq:composition_rate_bound} and integrating from $\Ye(\Rout)=0.5$ then gives
\begin{equation}\label{eq:Ye_analytic}
 \Ye^{\min}(\Runb)
 \gtrsim
 \frac{1}{2\sqrt{1+4\mathcal B_{\rm env}}}.
\end{equation}
Therefore, the condition $\mathcal B_{\rm env}<3/4$ is sufficient to exclude reaching the adopted threshold $\Ye^{\rm min}=0.25$. Finally, evaluating $\mathcal B_{\rm env}$ for the reference profiles $H_0$ and $\zeta_0$ from $\Rout=100\,r_g$ to $\Runb=10\,r_g$ gives
\begin{equation}
\label{eq:Benv_parameters}
\begin{split}
 \mathcal B_{\rm env}&\lesssim{}0.51
 \left(\frac{P_j}{5\times10^{51}\,{\rm erg\,s}^{-1}}\right)^2
 \left(\frac{\eta_a(a)}{\eta_a(0.3)}\right)^{-2}
 \left(\frac{\phibh}{20}\right)^{-4}\\\times
 &\left(\frac{\Risco(a)}{\Risco(a=0.3)}\right)^{-3}
 \left(\frac{\mbh}{3\,\Msun}\right)^{-3}\left(\frac{H}{H_0}\right)^{-5}
 \left(\frac{\zeta}{\zeta_0}\right)^{-3},
\end{split}
\end{equation}
where $H/H_0$ and $\zeta/\zeta_0$ are radially constant multipliers of the complete reference profiles, and the spin dependence follows $\eta_a(a)$ in Eq.~\eqref{eq:eta_a}. Fig.~\ref{fig:numerical}(d) shows that the analytic bound (dashed line) provides a reasonable estimate to the $\Ye$ in the collapsar simulations (dark solid line) after jet launching, $t\gtrsim 5\times 10^{3}\,t_g$ (depending on spin). At early times, before jet launching, the neutrinosphere reaches $R_{\nu_{\rm e}}\approx10\,r_g$ (Fig.~\ref{fig:neutrinosphere}), such that the transparent rates of Eq.~\eqref{eq:net_neutronization} do not apply, and $\Ye^{\min} > Y_{\rm e;\,avg}$. However, before jet launching, the gas remains bound, so this transitional phase has no effect on the ejecta composition.

For the adopted GRB power ceiling and fiducial BH parameters, collapsar disks cannot reach $Y_e\lesssim0.25$ at $R_{\rm unb}$. We stress that our estimates correspond to the most favorable corner of the collapsar parameter space: the upper end of the GRB peak luminosity distribution, low BH mass and spin, substantial sub-MAD jet launching, an unbinding radius of $10\,r_g$ at every flux, and negligible ram pressure on disk outflows. Additionally, our estimates are evaluated at the midplane, which is favorable for low-$\Ye$ ejecta, since winds are launched from the disk surface or high-latitude regions, where the density and degeneracy are lower and weak interactions drive $n/p$ toward unity \citep{Metzger2008}. This optimistic analysis is further demonstrated in Fig.~\ref{fig:numerical}(d), which shows that the analytic bound, consistent with the full numerical estimate (Fig.~\ref{fig:dependencies}), vastly underestimates the $\Ye$ values found in full GRMHD simulations.

\subsection{Gradual magnetization}
A possible loophole is temporal separation: the disk may first neutronize while weakly magnetized, and only later accumulate enough coherent flux to eject the gas that was neutronized earlier. This requires the field to reach ejection strength before that material is accreted, $t_{\rm grow}\lesssim t_{\rm res}$, where $t_{\rm res}$ is its remaining residence time.

The local weak-interaction relaxation time is
\begin{equation}
t_{\rm eq}^{-1}=\left.\frac{\partial\Gamma}{\partial\Ye}\right|_{\rho,T,\Ye=Y_{\rm e;\,equilibrium}}.
\label{eq:weak_relaxation}
\end{equation}
In the hot, nondegenerate limit, this reduces to
\begin{equation}
t_{\rm eq}
\simeq
\left(\lambda_{e^-p}+\lambda_{e^+n}\right)^{-1}
\simeq
10\,{\rm ms}
\left(\frac{T}{3\times10^{10}\,{\rm K}}\right)^{-5},
\end{equation}
using the capture-rate approximation of \citet[][Eqs.~43--44]{Beloborodov2003}. If the magnetized disk has $Y_{\rm e;\,equilibrium}\gtrsim0.4$ at $R_{\rm unb}$ [Fig.~\ref{fig:numerical}(d)], then at the launching densities along the GRB ceiling, $\rho\simeq(0.8$–$4.8)\times10^9\,{\rm g\,cm^{-3}}$, $kT\gtrsim2\,$MeV and Eq.~(\ref{eq:weak_relaxation}) gives $t_{\rm eq}\lesssim25\,{\rm ms}$. Since $t_{\rm eq}$ is shorter than the magnetic flux advection timescale $\Delta t_{\rm adv}\simeq70\,{\rm ms}\,(M_{\rm BH}/3\,M_\odot)$ \citep{Jacquemin-Ide2024b}, even fully neutronized gas relaxes to $\Ye\gtrsim0.38$ before ejection. Under these conditions, earlier neutronization cannot preserve sufficiently neutron-rich gas until ejection.

\section{Implications}
\label{sec:others}

\begin{figure*}
    \centering
    \includegraphics[width=1\linewidth]{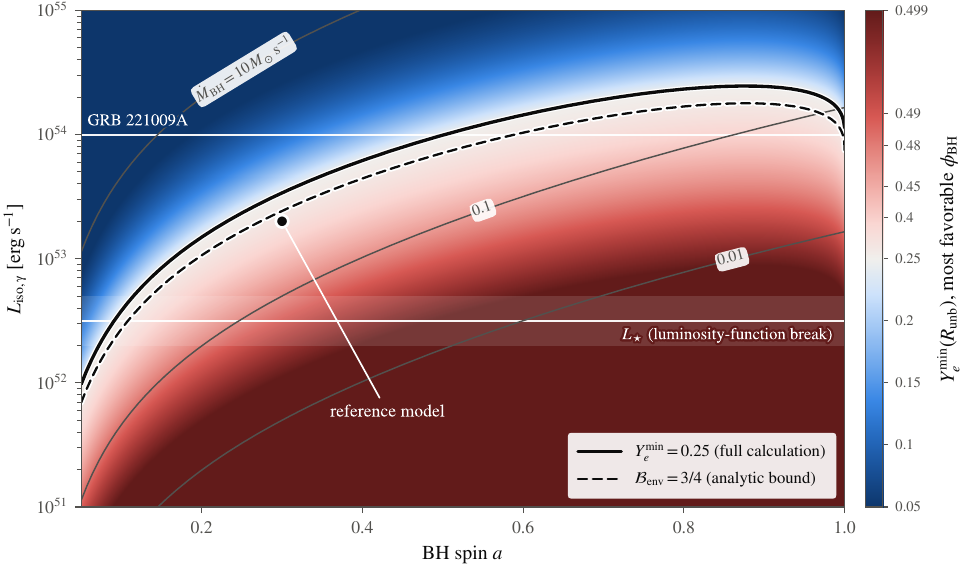}
    \caption{Most favorable composition bound as a function of black hole spin $a$ and isotropic-equivalent $\gamma$-ray luminosity $L_{\rm iso}$. Colors show $\Ye^{\min}(\Runb)$ derived from Eq.~\eqref{eq:neutronization_bound} with $\eta_a(a)$ of Eq.~\eqref{eq:eta_a}, assuming $\mbh=3\,\Msun$, $\theta_j=0.1$, and $\eta_\gamma=0.2$, and minimizing over $20\le\phibh\le50$ (minimum is reached at $\phibh=20$). The circle marks the ceiling reference model ($a=0.3$, $P_j=5\times10^{51}\,{\rm erg\,s^{-1}}$), for which $\Ye^{\min}\approx0.33$. The solid curve marks $\Ye^{\min}=0.25$ from the full numerical calculation and the dashed curve $\mathcal B_{\rm env}=3/4$, where the analytic bound of Eq.~\eqref{eq:Ye_analytic} reaches $\Ye=0.25$. Thin gray curves show $\mdotbh(\phibh{=}20)=0.01$, 0.1, and $10\,\Msun\,{\rm s^{-1}}$. Horizontal lines mark observed isotropic-equivalent peak luminosities: the break of the long-GRB luminosity function, $L_\star=10^{52.5\pm0.2}\,{\rm erg\,s^{-1}}$ (line and shaded uncertainty; \citealt{Wanderman2010}) and GRB~221009A ($9.9\times10^{53}\,{\rm erg\,s^{-1}}$; \citealt{Lesage2023}). Peak luminosities overestimate the sustained jet power and hence $\mdotbh$: GRB~221009A radiated $E_{\gamma,\rm iso}=1.0\times10^{55}$\,erg over more than 600\,s at $z=0.151$ \citep{Lesage2023}, corresponding to a mean luminosity of $\lesssim2\times10^{52}\,{\rm erg\,s^{-1}}$. The color scale is linear in $\Ye^{\min}$ below 0.25 (blue) and logarithmic in $0.5-\Ye^{\min}$ above it (red).}
    \label{fig:feasibility}
\end{figure*}

The constraint on high $\Ye^{\rm min}$ at $\Runb$ can be evaded by changing the origin of the low-$\Ye$ material or the difficulty of ejecting it. Compact-object mergers differ from collapsars in both respects. Some ejecta are launched dynamically from NS matter and therefore do not need to be neutronized in the disk. Post-merger disk winds are also launched into a comparatively clean, highly aspherical environment, rather than against the quasi-spherical ram pressure and baryon loading of a collapsing stellar envelope. Merger disks can therefore unbind material with much weaker magnetic flux, or plausibly through thermal and viscous driving \citep{Fernandez2013,Siegel2017}. Indeed, numerical simulations find that even weakly magnetized merger engines with $\phibh\lesssim1$ can launch polar outflows into the cleaner surrounding medium \citep[e.g.,][]{Gottlieb2023c}.

A different way to evade the collapsar bound is to relax the intrinsic engine power constraints. Weaker engines, including weak jets that fail to break out, only exacerbate the problem: at fixed magnetic flux, reducing $P_j$ lowers the allowed accretion rate and density, further suppressing neutronization. However, if the engine power is much larger than the observed long GRB scale, then a MAD-level outflow can, in principle, satisfy both the launching and neutronization requirements. This is the regime found by \citet{Issa2025}, where extremely high accretion rates allow the disk to remain neutrino cooled while still powering strong magnetic outflows. To visualize the dependency of $\Ye^{\rm min}$ on the GRB properties, Fig.~\ref{fig:feasibility} shows the jet power and BH spin dependencies of Eq.~\eqref{eq:Benv_parameters} with the full $\eta_a(a)$ of Eq.~\eqref{eq:eta_a} (dashed line) and of the full numerical calculation (solid line). It shows that under optimistic assumptions, reaching $\Ye^{\rm min}\approx 0.25$ requires the upper end of the GRB luminosity distribution and the lowest BH spin for which the jet can still dominate over disk outflows \citep{McKinney2012,Tchekhovskoy2012}.

This leaves two possible transient classes in which collapsar-like $r$-process ejecta may still be produced. The first is the extreme high-luminosity tail of the long-GRB population, including events such as GRB~221009A \citep{Burns2023,Lesage2023}. However, a large isotropic-equivalent luminosity allows low-$\Ye$ ejecta only if it originates in an unusually high accretion rate. Since the isotropic-equivalent jet power depends strongly on BH spin and the jet opening angle, the brightest GRBs are more naturally produced, at least in part, by higher spins and/or narrower angles rather than by proportionally larger accretion rates. In such cases, their large apparent luminosities do not allow low-$\Ye$ ejection as their increased luminosity is not translated to increased $\mdotbh$.

Even if some extremely luminous GRBs are powered by larger accretion rates, their volumetric rate is likely too small to represent the generic missing source of prompt $r$-process enrichment in low-metallicity environments, unless each event ejects an exceptionally large mass of heavy $r$-process material. Observationally, despite its extreme luminosity, the supernova of the nearby GRB~221009A showed no evidence for an additional $r$-process-powered component \citep{Blanchard2024}. More broadly, optical/near-infrared searches of SNe Ic-BL and GRB--SNe for the $r$-process signatures predicted by \citet{Barnes&Metzger2022} have found no compelling evidence for an $r$-process component \citep{Anand2024,Rastinejad2024,Blanchard2024}. Thus, the high-luminosity GRB tail may remain a rare exception, but it does not rescue collapsars as generic heavy-$r$-process factories. Stellar abundances also imply that collapsars are difficult to reconcile with metal-poor stars, which require early $r$-process production to be separated from iron production \citep{Macias2019}, and metallicity-suppressed collapsars cannot be the dominant $r$-process source in the Milky Way \citep{Fraser2022}. However, \citet{Brauer2021} argued that collapsar yields can reproduce the [Eu/Fe] scatter in metal-poor stars, such that abundances alone do not settle the question.

The second possibility is direct-horizon collapsars. These events may be rare at low redshift but more common in the early Universe. In metal-poor environments, reduced line-driven mass loss allows very massive stars to retain larger helium cores and greater angular momentum, increasing the likelihood of direct BH formation with much larger disk masses and accretion rates \citep{Siegel2022,Gottlieb2025}, allowing them to overcome the constraints imposed for collapsars. The full neutrino-transport calculation in GRMHD simulations of direct-horizon collapsars will be presented in Issa et al., in prep.

\section{Conclusions}\label{sec:conclusions}

We have examined whether collapsar accretion disks can produce heavy $r$-process ejecta through disk outflows. The basic requirement is that gas must become neutron-rich at the radius $\Runb$ outside which magnetic stresses can do enough work per loaded baryon to unbind the gas. We expressed this requirement as $\Ye^{\rm min}(\Runb) \lesssim 0.25$. For the adopted GRB engine parameters, this condition is not satisfied at any point along the magnetic-flux sequence.

In the weak-flux limit, the disk can remain dense, cool efficiently, and neutronize, but the effective unbinding radius lies far outside the disk. Any hydrodynamic wind is limited by the small residual advected thermal-energy fraction that cannot drive outflows against ongoing stellar fallback. Coherent magnetic stresses enable ejection but also shorten the inflow time and thicken the disk relative to a viscous flow, thereby inhibiting neutronization. Our calculation considers favorable conditions for neutronization: the jet power lies at the upper end of the bulk peak luminosity of the long-GRB population, sub-MAD jet launching is efficient, the BH mass is $3\,\Msun$, the temperature maximizes the net neutronization rate at every radius, and the ejecta carry the composition of the dense midplane. Even under these favorable conditions, the gas reaches $\Runb=10\,r_g$ with $\Ye^{\min}\gtrsim0.33$ for every flux capable of ejection, from minimum ejecting flux through the MAD regime. In particular, $\Ye^{\min}\approx0.33$ is obtained for the minimum flux needed for sustaining jet launching. As jet launching drives the flow toward MAD, that phase is likely short-lived.

The obstruction identified here applies specifically to disk-wind nucleosynthesis in collapsars. It does not apply to compact-object mergers, not simply because some of their ejecta begin as neutron-rich NS matter, but because the ejection problem is fundamentally different. Merger remnants are not embedded in a quasi-spherical stellar fallback flow, so their winds encounter a much lower ram-pressure and baryon-loading barrier. In addition, part of the ejecta can be launched dynamically without requiring a disk wind to neutronize and unbind the same gas parcel. The collapsar bound can instead be evaded only by moving to more extreme high-power engines. Direct-horizon collapsars or other rare systems with much larger accretion rates may remain neutrino cooled despite strong magnetic stresses, as their engine power is not constrained by GRB observations.

\begin{acknowledgments}
\renewcommand{\makeLineNumber}{}
This research used resources of the Argonne Leadership Computing Facility, a U.S. Department of Energy (DOE) Office of Science user facility at Argonne National Laboratory and is based on research supported by the U.S. DOE Office of Science-Advanced Scientific Computing Research Program, under Contract No. DE-AC02-06CH11357 (NeutronStarRemnants project). JJ acknowledge support from NASA grants 80NSSC22K0826, 80NSSC24K0941, and 80NSSC24K1174. J.J. acknowledges support by NSF grants AST-2307983.
\end{acknowledgments}

\bibliographystyle{aasjournal}
\bibliography{refs}

\appendix
\section{Neutronization along inflow trajectories}
\label{app:trajectories}

We follow Eq.~\eqref{eq:neutronization_bound} along individual inflow trajectories to show where the neutronization accumulates, which temperatures the optimization selects, and whether the free-nucleon, neutrino-transparent rates of \S\ref{sec:mad} are appropriate along those paths. We consider the GRB ceiling with $\mdotbh(\phibh)$ from Eq.~\eqref{eq:Mdotbh}. Each trajectory starts at $\Rout=100\,r_g$ with $\Ye=0.5$ and is integrated inward to $\Runb=10\,r_g$ with Eq.~\eqref{eq:neutronization_bound}. The density and inflow speed follow from Eqs.~\eqref{eq:magnetic_inflow} and \eqref{eq:rho}, with numerical fits to $H_0(r)$ and $\zeta_0(r)$, $\mbh=3\,\Msun$, and $\dMd=\mdotbh$. The optimized rate $\Gamma_\star(\rho,\Ye)$ is tabulated from the full capture integrals, maximizing over temperature. The temperature $T_\star(r)$ is the maximizing temperature at the local $\rho(r)$ and $\Ye^{\min}(r)$.

\begin{figure}
    \centering
    \includegraphics[width=1\linewidth]{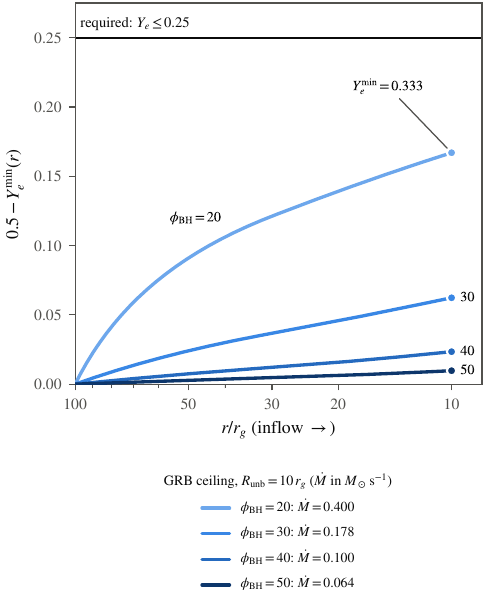}
    \caption{Temperature-optimized composition along the inflow. Curves show $0.5-\Ye^{\min}(r)$ from Eq.~\eqref{eq:neutronization_bound}, integrated inward from $\Rout=100\,r_g$ with $\Ye=0.5$, for points on the GRB ceiling with $\phibh=20$, 30, 40, and 50. Circles mark the common adopted launching radius, $R_{\rm unb}=10\,r_g$, where $\Ye^{\min}=0.33,0.44,0.48,$ and $0.49$ for $\phi_{\rm BH}=20,30,40,$ and $50$, respectively.}
    \label{fig:composition_paths}
\end{figure}

Fig.~\ref{fig:composition_paths} shows $\Delta\Ye(r)\equiv0.5-\Ye^{\min}(r)$. Along the ceiling the final values are $\Delta\Ye=$0.01--0.17 ($\Ye^{\min}=$0.33--0.49). The processing decreases monotonically from $\phibh=20$ to 50, because the ceiling accretion rate scales as $\dot M_{\rm BH,max}\propto\phibh^{-2}$, lowering the density along the same $100\,r_g\rightarrow10\,r_g$ inflow trajectory.

\subsection{Dependence on the outer radius}
\label{app:rout}

Using the rate envelope $\Gamma_{\rm env}\propto\rho^2$, the maximum neutronization per logarithmic radius scales as $\zeta_0^{-3}(H_0/r)^{-5}r^{-3/2}$. With $\zeta_0\propto r^{-1}$, this becomes $\propto r^{3/2}(H_0/r)^{-5}$, which rises outward wherever $H_0/r$ grows more slowly than $r^{0.3}$. It therefore falls as $\sim r^{-0.4}$ between $10$ and $30\,r_g$, but rises as $\sim r^{0.4}$ between $30$ and $100\,r_g$. As a result, the free-nucleon bound does not converge with $\Rout$: extrapolating the profiles to $1000\,r_g$ gives $\Ye^{\min}=0.26$ at $\phibh=20$. However, outside $100\,r_g$ an increasing fraction of the nucleons is bound in $\alpha$ particles. We therefore repeat the integration from $3000\,r_g$ assuming nuclear statistical equilibrium of neutrons, protons, and $\alpha$ particles, and maximize the net rate of Eq.~\eqref{eq:app_alpha_rate} over temperature. The result converges, since starting at $10^4\,r_g$ gives the same values: $\Ye^{\min}=0.31$ at $\phibh=20$, $0.02$ below the value for $\Rout=100\,r_g$, while for $\phibh\gtrsim30$ the change is $\lesssim0.004$. The approximation $S=1/2$, which holds at $4\,\Risco\approx20\,r_g$ and sets $\Rout$ in \S\ref{sec:mad}, gives $\Ye^{\min}=0.33$. This likely overestimates the heating in our coherent-torque picture, in which part of the released energy is carried away by the wind \citep{ferreira_magnetized_1995}.

\begin{figure*}
    \centering
    \includegraphics[width=1\linewidth]{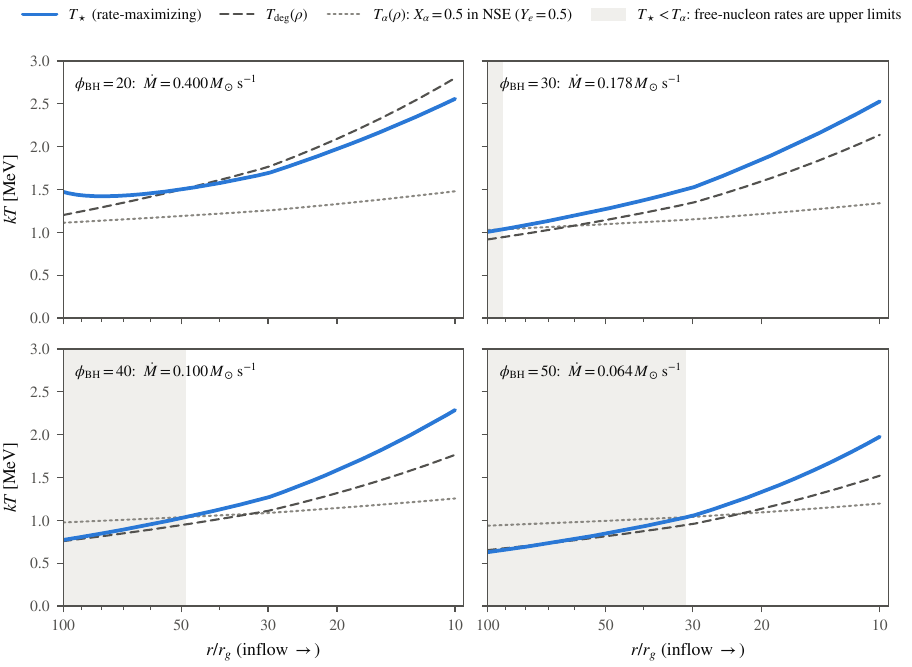}
    \caption{Rate-maximizing temperature along the four ceiling trajectories of Fig.~\ref{fig:composition_paths}; each panel lists the maximum $\mdotbh$. Solid curves show $T_\star(r)$ at the local density $\rho(r)$ and composition $\Ye^{\min}(r)$ [Eq.~\eqref{eq:neutronization_bound}]. Dashed curves show the degeneracy temperature $T_{\rm deg}[\rho(r)]$ [Eq.~\eqref{eq:Tdeg}], and dotted curves show $T_\alpha[\rho(r)]$, the temperature at which $\alpha$ particles carry half the mass in nuclear statistical equilibrium of neutrons, protons, and $\alpha$ particles with $\Ye=0.5$ [Eq.~\eqref{eq:app_talpha}]. In the shaded ranges, $T_\star<T_\alpha$ and the free-nucleon rates overestimate the neutronization.}
    \label{fig:temperature_paths}
\end{figure*}

\subsection{Rate-maximizing temperature}
\label{app:tstar}

The rate-maximizing temperatures in Fig.~\ref{fig:temperature_paths} range from $kT_\star=0.63$\,MeV at $100\,r_g$ to 2.6\,MeV at $10\,r_g$ along the ceiling. Their scaling follows from the mildly degenerate expansion of Eq.~\eqref{eq:rate_expansion}, whose maximum lies at
\begin{equation}
\label{eq:app_tstar}
kT_\star\simeq2.3\,{\rm MeV}\left(\frac{\rho}{10^9\,{\rm g\,cm^{-3}}}\right)^{1/2}=\frac{kT_n}{\sqrt2},
\end{equation}
where $kT_n\simeq3.3\,{\rm MeV}\,(\rho/10^9\,{\rm g\,cm^{-3}})^{1/2}$ is the temperature above which the equilibrium composition of transparent, free-nucleon matter becomes proton-rich \citep[][Eq.~14]{Beloborodov2003}. The full integrals give $T_\star/T_n=0.71$--0.77 for $10^8\le\rho\le10^{10}\,{\rm g\,cm^{-3}}$ at $\Ye=0.5$. At higher temperature, the matter is less degenerate and positron captures increasingly offset electron captures, reversing the net rate above $T_n$. At temperatures lower
than the degeneracy temperature (dashed curves),
\begin{equation}
\label{eq:Tdeg}
kT_{\rm deg}\simeq1.7\,{\rm MeV}\left(\frac{\rho}{10^9\,{\rm g\,cm^{-3}}}\right)^{1/3},
\end{equation}
electrons are degenerate \citep[][Eq.~4]{Beloborodov2003}, but fewer electrons exceed the capture threshold. $T_\star$ also falls as the matter neutronizes, since a lower $\Ye$ reduces the degeneracy. Therefore, the optimal state lies near or above $T_{\rm deg}$ along all paths.

\subsection{$\alpha$ particles}
\label{app:alpha}

The capture rates assume free nucleons. In nuclear statistical equilibrium of neutrons, protons, and $\alpha$ particles with $\Ye=0.5$, half of the mass is in $\alpha$ particles when
\citep[e.g.,][]{Qian1996}
\begin{equation}
\label{eq:app_talpha}
\left(\frac{\rho}{m_u}\right)^{3}\left(\frac{2\pi\hbar^2}{m_u\,kT_\alpha}\right)^{9/2}\exp\!\left(\frac{28.3\,{\rm MeV}}{kT_\alpha}\right)=64.
\end{equation}
At $\Ye=0.5$, the optimized temperature falls below $T_\alpha$ for $\rho\lesssim1.8\times10^8\,{\rm g\,cm^{-3}}$ (shaded in Fig.~\ref{fig:temperature_paths}), consistent with the region that \citet[][Fig.~1]{Beloborodov2003} marks as dominated by composite nuclei. Nucleons bound in $\alpha$ particles do not capture electrons or positrons, such that at fixed $\rho$, $T$, and $\Ye$ the net rate of Eq.~\eqref{eq:net_neutronization} becomes
\begin{equation}
\label{eq:app_alpha_rate}
\Gamma_\alpha=\Gamma-\frac{X_\alpha}{2}\left(\lambda_{e^-p}-\lambda_{e^+n}\right).
\end{equation}
Wherever $\lambda_{e^-p}>\lambda_{e^+n}$, which holds along all the trajectories considered here, $\alpha$ formation lowers the net rate. Where $\lambda_{e^-p}<\lambda_{e^+n}$, both $\Gamma$ and $\Gamma_\alpha$ are negative for $\Ye\le0.5$. Therefore, $\max_T\Gamma_\alpha\le\Gamma_\star$, and $\Ye^{\min}$ remains a lower limit. Heavier nuclei, which appear at still lower temperatures, typically have much smaller capture rates per nucleon than free protons at these densities \citep{Langanke2003}. At $T_\star$ the $\alpha$ mass fraction reaches 0.99 and 0.93 in the outer parts of the $\phibh=50$ and 40 paths, where the $\alpha$-dominated ranges contribute 46\% and 33\% of the final $\Delta\Ye$. The $T_\star(\phibh=30)$ path falls below $T_\alpha$ only near $\sim100\,r_g$, where it accrues $\sim6\%$ of its $\Delta\Ye$. Along the $\phibh=20$ path, $X_\alpha\lesssim0.02$, which lowers the net rate by about $3\%$. However, the rate-maximizing temperature then shifts slightly upward, such that including $\alpha$ particles and re-maximizing over temperature raises $\Ye^{\min}$ by only 0.001--0.004 along the ceiling.

\subsection{Neutrino absorption}
\label{app:neutrinos}

\begin{figure}
    \centering
    \includegraphics[width=1\linewidth]{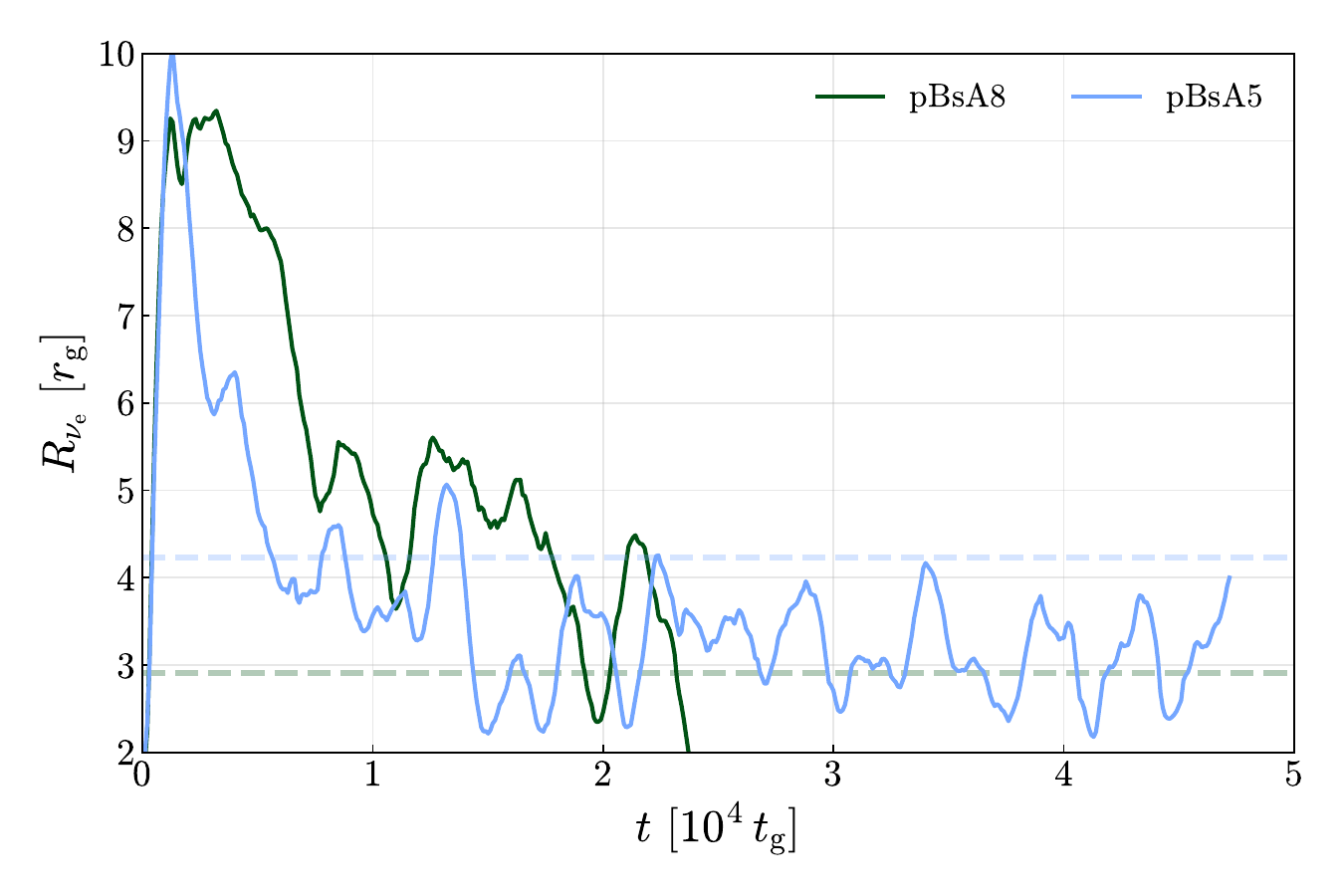}
    \caption{
    Time evolution of the electron neutrinosphere radius, $R_{\nu_{\rm e}}$, at which the integrated effective optical depth reaches $\tau_{\rm eff} = \int_{R_{\nu_{\rm e}}}^{\infty} dr \sqrt{\kappa_a (\kappa_a + \kappa_s)} = 2/3$, where $\kappa_{a,s}$ are absorption/scattering opacities respectively. $R_{\rm ISCO}(a)$ are shown with dashed horizontal lines. The neutrinosphere remains inside $\approx10\,r_g$, supporting the transparent-matter approximation over the radial domain used in \S\ref{sec:mad}.
    }
    \label{fig:neutrinosphere}
\end{figure}

The GRMHD simulation data in Fig.~\ref{fig:neutrinosphere} show that the electron neutrinosphere recedes to $R_{\nu_{\rm e}}\lesssim 8\,r_g$ once the jet is launched, and to $R_{\nu_{\rm e}}\lesssim 6\,r_g$ after $t\approx10^4\,t_g$, such that the gas at $r\geq10\,r_g$, where we integrate Eq.~\eqref{eq:neutronization_bound}, is transparent. In our model, the capture spectra at $T_\star$, $\langle E_\nu^2\rangle\approx(29$--$30)(kT_\star)^2$, and $\sigma_\nu\approx9\times10^{-44}(E_\nu/{\rm MeV})^2\,{\rm cm^2}$ per nucleon \citep[][Eqs.~54, 59]{Beloborodov2003} give an optical depth across the flow at $R_{\rm unb}$ of $\tau_\nu\approx\sigma_\nu(\rho/m_u)H\lesssim0.05$ along the ceiling and $\leq0.085$ along the $\Ye^{\rm min}=0.25$ boundary of Fig.~\ref{fig:dependencies}.

However, a transparent flow can still absorb neutrinos emitted by the rest of the disk \citep{Surman2004,Surman2005}. Integrating the $\nu_e$ and $\bar\nu_e$ emission of the whole flow at $T_\star$, extending each ceiling path inward to $5\,r_g$ with $H/r=0.23$ and $\zeta=1.34$, we find absorption rates at $R_{\rm unb}$ of $0.2$--$7\,{\rm s^{-1}}$ for $\nu_e$ and $0.15$--$2.8\,{\rm s^{-1}}$ for $\bar\nu_e$, compared with $\Gamma_\star=3.9$--$25\,{\rm s^{-1}}$, such that each nucleon absorbs $\lesssim0.05$ neutrinos before it is launched. Since electron captures dominate at $T_\star$, the disk emits $1.9$--$2.5$ times more $\nu_e$ than $\bar\nu_e$, driving $\Ye$ toward $0.6$--$0.7$ and raising $\Ye^{\rm min}$ by $\leq0.008$. A hotter inner disk, at the maximum temperature of \citet[][Eq.~41]{Beloborodov2003}, drives $\Ye$ toward $\sim 0.48$, raising $\Ye^{\rm min}$ by $0.07$ for $\phi_{\rm BH}=20$, while leaving it largely unchanged for $\phi_{\rm BH}=30$--$50$. Therefore, neglecting neutrino absorption keeps the minimum along the ceiling a lower limit.

\end{document}